\documentclass[twocolumn,preprintnumbers,superscriptaddress,nofootinbib,aps,prd,floatfix]{revtex4-2}
\pdfoutput=1
\usepackage{enumerate}
\usepackage{gensymb}
\usepackage{amsmath,amssymb}
\usepackage{mathrsfs}
\usepackage{graphicx}
\usepackage{slashed}
\usepackage{xspace,slashed}
\usepackage{hyperref}
\usepackage[dvipsnames]{xcolor}
\hypersetup{colorlinks=true, citecolor=Blue, urlcolor=Blue, linkcolor=Blue}
\usepackage[normalem]{ulem}
\usepackage{subfigure,orcidlink}
\usepackage[autostyle]{csquotes}
\usepackage{multirow,array}
\usepackage{float}
\usepackage{tabularray}
\UseTblrLibrary{booktabs}
\usepackage[absolute,overlay]{textpos} 

\begin{document}

\title{Resurrecting Electroweak Dark Matter via Type-II Seesaw \\in light of recent LZ Event}

\author{Partha Kumar Paul \orcidlink{0000-0002-9107-5635}}
\email{ph22resch11012@iith.ac.in}
\affiliation{Department of Physics, Indian Institute of Technology Hyderabad, Kandi, Telangana-502285, India.}

\author{Sujit Kumar Sahoo \orcidlink{0000-0002-9014-933X}}
\email{sujitks@imsc.res.in}
\affiliation{Department of Physics, Indian Institute of Technology Hyderabad, Kandi, Telangana-502285, India.}
\affiliation{Institute of Mathematical Sciences (IMSc), Chennai 600113, India.}

\author{Narendra Sahu \orcidlink{0000-0002-9675-0484}}
\email{nsahu@phy.iith.ac.in}
\affiliation{Department of Physics, Indian Institute of Technology Hyderabad, Kandi, Telangana-502285, India.}

\author{Shashwat Sharma{\orcidlink{0009-0002-2266-0467}}}
\email{ph23resch11016@iith.ac.in}
\affiliation{Department of Physics, Indian Institute of Technology Hyderabad, Kandi, Telangana-502285, India.}
	
\date{\today}
\begin{abstract}
The recent observation of the high-energy nuclear recoil event LZ230616 with a recoil energy around $248$ keV provides an interesting possibility to probe inelastic doublet dark matter (iDM). However, the electroweak doublet DM scenarios face strong constraints from solar DM capture for DM masses around the TeV scale which give rise to correct thermal relic density. In this work, we revive the inelastic doublet DM (iDM) scenario via type-II seesaw. In particular, we consider an inert lepton doublet (ILD) DM. While the minimal ILD scenario gives the correct relic abundance around the TeV scale, it is strongly constrained by direct detection and solar DM capture. In the type-II seesaw, the presence of the scalar triplet generates the required mass splitting and provides additional annihilation channels, allowing the correct relic abundance for much larger DM masses. The larger DM mass also helps evade the solar capture and indirect detection constraints. Additionally, the scalar triplet gives sub-eV neutrino masses required by the oscillation data.
\end{abstract}

\maketitle	
\noindent
\section{Introduction}
\label{sec:intro}

Recently, the LZ experiment reported a nuclear recoil event with reconstructed energy $E_R=248\pm23_{\rm stat}\pm23_{\rm sys} ~{\rm keV}$, with an exposure of $2.84$ tonne-years and an extended nuclear recoil energy range from 5.4 keV to about $270$ keV \cite{LZ:2026axp}. The event lies in an energy region where the expected background is small, providing an interesting opportunity to investigate DM interactions at high recoil energies. In the standard elastic DM case, it gives a spectrum that increases towards lower recoil energies. Therefore, explaining an isolated event at such a high recoil energy while avoiding an excess at lower energies is difficult within the usual elastic WIMP framework.

Inelastic dark matter (iDM) is one of the interesting possibilities where the dark sector contains two states, $\chi_1$ and $\chi_2$, with a small mass splitting $\delta=m_{\chi_2}-m_{\chi_1}$ \cite{Tucker-Smith:2001myb,Tucker-Smith:2004mxa,Cui:2009xq,Arina:2012aj,Borah:2020smw,Cho:2024lhp}. The presence of the mass splitting changes the scattering kinematics and thus introduces a nonzero energy cost for the transition $\chi_1 N\rightarrow \chi_2 N$. As a result, the recoil spectrum can be concentrated in a relatively narrow region of recoil energy, providing a natural way to obtain high-energy nuclear recoils without generating a large low-energy signal. 

A well-motivated realization of such a framework arises in Higgsino-like electroweak doublet DM scenarios, where the correct DM relic abundance is obtained for a DM mass of around $1.1$ TeV \cite{Hall:1997ah,Martin:1997ns}. Such a scenario can explain the observed LZ230616 event through inelastic scattering with a mass splitting of approximately $350$ keV. However, Higgsino-like electroweak doublet DM \cite{Fan:2026kxx,Freese:2026sga,Wu:2026nhi,Yin:2026jnn,Rodd:2026tyn,Smirnov:2026aqk,Nomura:2026qyq} with a mass around $1$ TeV is strongly constrained by the solar capture bound \cite{Pospelov:2026ewn,Bose:2026ndd,Nguyen:2026lui,Langhoff:2026ujr}.

Another realization of electroweak doublet DM is the inert lepton doublet (ILD) scenario~\cite{Bhattacharya:2018fus}. In the minimal ILD scenario, the neutral component of the doublet serves as the DM candidate, with the correct relic abundance typically obtained for a mass around the TeV scale~\cite{Bhattacharya:2018fus}. However, the same electroweak interactions responsible for DM scattering and annihilation also lead to strong constraints from direct and indirect detection experiments, as well as from DM capture in the Sun~\cite{Bhattacharya:2018fus,Pospelov:2026ewn,Bose:2026ndd,Nguyen:2026lui}. Several recent studies have explored possible
explanations for the observed LZ event \cite{Du:2026guj, McCabe:2026crm, Unwin:2026rdp,  Lou:2026idn, Su:2026rwz, DiMauro:2026ldr, Yamashita:2026ump, Chattopadhyay:2026ryw, deLima:2026shq, Visinelli:2026kgt,Borah:2026zwf,Mahapatra:2026glu,Das:2026uyy,Borah:2026ris,Bandyopadhyay:2026gjw,Barman:2026omh,Lee:2026jxl,Nagata:2026pbj,Arcadi:2026kev,Fan:2026hzw,He:2026hqz,Kumar:2026lgi,Lee:2026jxl,Baer:2026fpy,Okada:2026upm,He:2026idw,Uttayarat:2026isp,Palmisano:2026kuj,Lian:2026hpm,An:2026pkc}.

In this work, we consider an extension of the ILD scenario by introducing a scalar triplet \cite{Bhattacharya:2018fus,Arina:2012aj,Barman:2019tuo}. After electroweak symmetry breaking, the triplet obtains an induced vacuum expectation value (VEV). This splits the neutral component of the doublet into two states with a small mass splitting ($\delta$). The DM candidate can therefore naturally interact inelastically with nuclei, allowing the high energy LZ event to be studied within an iDM framework. An important feature of this construction is that the mass splitting is controlled by the triplet VEV rather than being introduced as an independent parameter. The role of the scalar triplet is three-fold: (i) it generates a small mass splitting between the neutral components of the doublet, (ii) it brings the overabundant DM relic density to the observed value even for large DM masses, and (iii) it generates Majorana neutrino masses through the type-II seesaw mechanism~\cite{Magg:1980ut,Lazarides:1980nt,Mohapatra:1979ia,Mohapatra:1980yp,Ma:1998dx}. It is also worth noting that the baryon asymmetry of the Universe can be addressed within this framework through leptogenesis \cite{Ma:1998dx,Hambye:2001eu,Hambye:2005tk}. This mechanism can apply to fourth generation vector-like neutrino DM \cite{Arina:2012aj}, as well as sneutrino DM in a supersymmetric scenario \cite{Chatterjee:2014vua}.

The paper is organized as follows. In Section~\ref{sec:LZevent}, we present a model-independent analysis of the LZ event within the framework of vector-mediated inelastic scattering; in Section~\ref{sec:model}, we provide a model realization and identify the relic-compatible parameter space that explains the LZ event while satisfying the solar capture bound. We conclude in Section~\ref{sec:conclusion}.

\section{inelastic DM and LZ event}\label{sec:LZevent}

We first examine the kinematic conditions required to explain the LZ event by assuming that it arises from inelastic DM scattering off a nucleus, $\chi_1 N \to \chi_2 N$. In this process, the incoming DM particle must have a minimum velocity to produce a nuclear recoil with energy $E_R$. This minimum velocity is given by \cite{Tucker-Smith:2001myb}
\begin{equation}
\label{eq:vmin}
v_{\rm min}=\sqrt{\frac{m_NE_R}{2\mu_N^2}}+\frac{\delta}{\sqrt{2E_Rm_N}},
\end{equation}
where $m_N$ denotes the nuclear mass. For the Xenon ($^{131}$Xe) target used in the LZ detector, we take $m_N\simeq122$ GeV, while $\mu_N$ is the reduced mass of the DM-nucleus system. When the scattering is elastic, i.e. $\delta\rightarrow0$, the second term vanishes and $v_{\rm min}$ increases with increasing recoil energy. For inelastic scattering, however, the final-state particle $\chi_2$ is heavier than the initial state $\chi_1$ by an amount $\delta$. Consequently, some additional energy is needed to produce $\chi_2$, which leads to the second term in $v_{\rm min}$. Since this contribution is proportional to $\delta/\sqrt{E_R}$, it becomes increasingly relevant at low recoil energies.

For a given mass splitting $\delta$, $v_{\rm min}$ reaches its lowest value at a particular recoil energy,
\begin{equation}
\label{eq:ERforvmin}
E_R=\frac{\mu_N}{m_N}\delta.
\end{equation}
This shows that the recoil energy corresponding to the most favorable inelastic kinematics is directly connected to the mass splitting. Its dependence on the DM mass and the nuclear target enters only through the reduced mass. Therefore, this expression provides a simple way to relate a relatively high recoil-energy event to the range of mass splittings that can produce it.

For the DM velocity distribution, we use the Standard Halo Model (SHM). In this framework, the DM velocity distribution in the Galactic frame is described by a Maxwell-Boltzmann distribution with a high-velocity cutoff \cite{Lewin:1995rx},
\begin{equation}
\label{eq:velocitydistn}
f_{\text{SHM}}^{\text{gal}}=k e^{-v^2/v_0^2}\Theta(v_{\rm esc}-v),
\end{equation}
where the normalization factor is determined by
\begin{equation}
k^{-1}=(\pi v_0^2)^{3/2}\left[\text{erf}\left(\frac{v_{esc}}{v_0}\right)-\frac{2}{\sqrt{\pi}}\frac{v_{esc}}{v_0}e^{-v_{esc}^2/v_{0}^2}\right].
\end{equation}
Here, $v_0=220$ km/s is the velocity dispersion and $v_{\rm esc}=540$ km/s is the Galactic escape velocity. The function $\Theta$ is the Heaviside step function, which imposes the escape-velocity cutoff.

The corresponding differential recoil rate in a direct-detection experiment is given by
\begin{equation}
\label{eq:dRdER}
\frac{dR}{dE_R}=\frac{\rho_{\text{DM}}}{m_\text{DM}}N_T\int_{v>v_{\rm min}}v f^{\text{gal}}(\vec{v}+\vec{v_e}(t))\frac{d\sigma}{dE_R}d^3v,
\end{equation}
where $\vec{v}$ is the DM velocity in the Earth frame and $\vec{v}_e(t)$ denotes the Earth's velocity relative to the Galactic frame. We use a local DM energy density of $\rho_{\text{DM}}\simeq0.4$ GeV/cm$^3$ and assume that SDDM makes up the full DM abundance. The quantity $N_T$ represents the number of target nuclei per unit detector mass.

For spin-independent DM scattering on a nucleus, the differential cross section can be written as
\begin{equation}
\label{eq:dsigmadER}
\frac{d\sigma}{dE_R}=\frac{\sigma_{n}}{v^2}\frac{m_N}{2\mu_n^2}\left(\frac{Zf_p+(A-Z)f_n}{f_n}\right)^2 F^2(E_R)\mathcal{G}_{\rm sm},
\end{equation}
where $\sigma_n$ is the DM-nucleon inelastic scattering cross section, and $\mu_n$ is the corresponding DM-nucleon reduced mass. The factor $\mathcal{G}_{\rm sm}$ accounts for the experimental energy smearing and is given by
\begin{eqnarray}
\mathcal{G}_{\rm sm}=\frac{1}{\sqrt{2\pi}\sigma_E}e^{-\frac{(E_{\rm obs}-E_R)^2}{2\sigma_E^2}},
\end{eqnarray}
with
\begin{equation}
E_{\rm obs}=248{\rm ~keV},~\sigma_E=1.46{~\rm keV}\sqrt{E_R/{\rm keV}}.
\end{equation}
Here, $A$ and $Z$ are the nuclear mass and atomic numbers, respectively, while $f_p$ and $f_n$ denote the effective couplings of DM to protons and neutrons. The finite size of the nucleus is taken into account through the nuclear form factor $F(E_R)$. We use the Helm form factor \cite{Engel:1991wq,Duda:2006uk},
\begin{equation}
\label{eq:Helmformfactor}
F^2(E_R)=\left(\frac{3j_1(qr_0)}{qr_0}e^{\frac{-(qs)^2}{2}}\right)^2,
\end{equation}
where $j_1(x)$ is the spherical Bessel function of the first kind and
\begin{equation}
q=\sqrt{2m_NE_R},\qquad
s\simeq0.9~{\rm fm},
\end{equation}
\begin{equation}
r_0=\sqrt{c^2+\frac{7}{3}\pi^2a^2-5s^2},
\end{equation}
with $a\simeq0.52$ fm and $c=1.23 A^{1/3}-0.6$.

Using the above expressions, we perform a model independent analysis to check which regions of the parameter space can account for the observed LZ event. We perform this study using the extended maximum-likelihood method \cite{BARLOW1990496}. The likelihood is defined in terms of the model parameters ${x_i}$ as
\begin{eqnarray}
{\mathcal L}({x_i}) & \equiv & \left[\prod_{i=1} P(x_i)\right]e^{-\mathcal{N}}, \\
& = & \left[\prod_{i=1}^{n_o} \left.\frac{d N({x_i})}{d E_R^\prime}\right|_{E_R^\prime=E_i}\right]e^{-\mathcal{N}({x_i})},\nonumber 
\end{eqnarray}
where the expected total number of events is
\begin{equation}
\mathcal{N}=\int_{E_R^{\rm min}}^{E_R^{\rm max}}\epsilon_{\rm LZ}(E_R^\prime)
\frac{dN({p})}{dE_R^\prime}dE_R^\prime,
\end{equation}
where $\epsilon_{\rm LZ}(E_R)$ is the detector efficiency. Thus, $\mathcal{N}$ gives the number of events predicted by a given set of model parameters within the recoil-energy interval $\left[E_R^{\rm min},E_R^{\rm max}\right]$. The differential event number is obtained from the recoil rate as
\begin{equation}
\frac{dN}{dE_R}=\frac{dR}{dE_R}\times\text{exposure}\times\epsilon_{\rm LZ}(E_R),
\end{equation}
where the LZ exposure is 2.84 tonne-year.

In our analysis, the relevant parameters are the DM mass $m_{\rm DM}$, the mass splitting $\delta$, and the inelastic DM-nucleon cross section $\sigma_{n}$. We consider one observed signal event, corresponding to $n_o=1$, with a recoil energy of
$E_i=248\pm23$(stat)$\pm23$(sys)$~{\rm keV}$. The recoil-energy window considered in the analysis is
\begin{equation}
E_R^{\rm min}=5.4~{\rm keV},\qquad
E_R^{\rm max}=269.9~{\rm keV},
\end{equation}
which covers the relevant energy range probed by the LZ experiment.

\begin{figure}[h]
    \centering
    \includegraphics[scale=0.4]{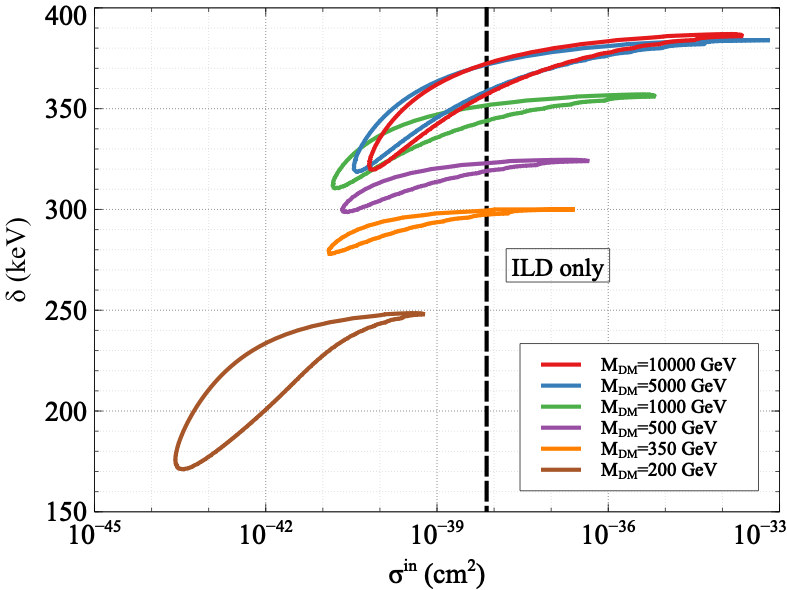}
    \caption{Model independent $1\sigma$ contours that could account for the LZ230616 event in the plane of $\delta$ vs $\sigma_{n}$ for different masses as shown in the figure.}
    \label{fig:LZ_ind_contour}
\end{figure}

We evaluate the likelihood $\mathcal{L}(\delta,\sigma_{n})$ by fixing the DM mass $M_{\text{DM}}$ over the parameter space of interest and determine its maximum numerically,
\begin{eqnarray}
\ln\mathcal{L}_{\rm max}
&=& \max_{\delta,\sigma_{n}}\ln\mathcal{L}(\delta,\sigma_{n})\\
&=& \max_{\delta,\sigma_{n}}
\left(-\mathcal{N}
+\ln\left(\left.\frac{dN}{dE_R}\right|_{E_R=E_i}\right)\right).\nonumber
\label{eq:maxlikelihood}
\end{eqnarray}
The corresponding values $(\delta^{\rm BF},\sigma_{n}^{\rm BF})$ define the best-fit point. We characterize the region preferred by the event using the likelihood-ratio statistic
\begin{equation}
\Delta\chi^2(\delta,\sigma_{n})
=-2\left[\ln\mathcal{L}(\delta,\sigma_{n})
-\ln\mathcal{L}_{\rm max}\right].
\label{eq:deltachi2}
\end{equation}
For two simultaneously varied parameters, we define the $1\sigma$ region using the standard likelihood-ratio criterion
\begin{equation}
\Delta\chi^2(\delta,\sigma_{n})\leq 2.30,
\label{eq:contourdef}
\end{equation}
corresponding to a $68.3\%$ confidence region under the usual $\chi^2$ approximation. The $1\sigma$ contours shown in Fig.~\ref{fig:LZ_ind_contour} are therefore obtained by numerically tracing the curve $\Delta\chi^2=2.30$ in the $(\delta,\sigma_{n})$ plane for different values of $M_{\text{DM}}$.

Following Fig.~\ref{fig:LZ_ind_contour}, we can see that the mass splitting required to fit the event increases together with the cross section along each contour. This reflects the kinematics of inelastic scattering: a larger
$\delta$ raises the minimum incoming DM speed $v_{\rm min}(E_R,\delta,M_\text{DM})$ required to up-scatter into the heavier state. Since the local DM velocity distribution falls steeply in this regime, a larger $\delta$ sharply suppresses the population of particles fast enough to scatter, and a correspondingly larger cross section is needed to compensate and still reproduce one expected event. 

The contours for the largest masses considered, $M_\text{DM} = 5$ and $10\,\mathrm{TeV}$, lie
almost on top of each other. This follows from the behavior of the DM--nucleus reduced mass $\mu_N = M_{\text{DM}} m_N/(M_{\text{DM}} + m_N)$, which enters the relation of the kinematic minimum of $v_{\rm min}(E_R)$ at fixed $\delta$, $E_R = (\mu_N/m_N)\,\delta$. As
$M_{\text{DM}} \to \infty$, $\mu_N/m_N \to 1$, and this ratio already lies within a few percent
of unity by $M_{\text{DM}} \sim 5\,\mathrm{TeV}$. The kinematic structure of the problem has
therefore already converged to its heavy-mass limit at these masses, so increasing
$M_{\text{DM}}$ further produces negligible additional change in the preferred
$(\delta,\sigma_{n})$ region.

\begin{figure}[h]
\centering
\includegraphics[width=1\linewidth]{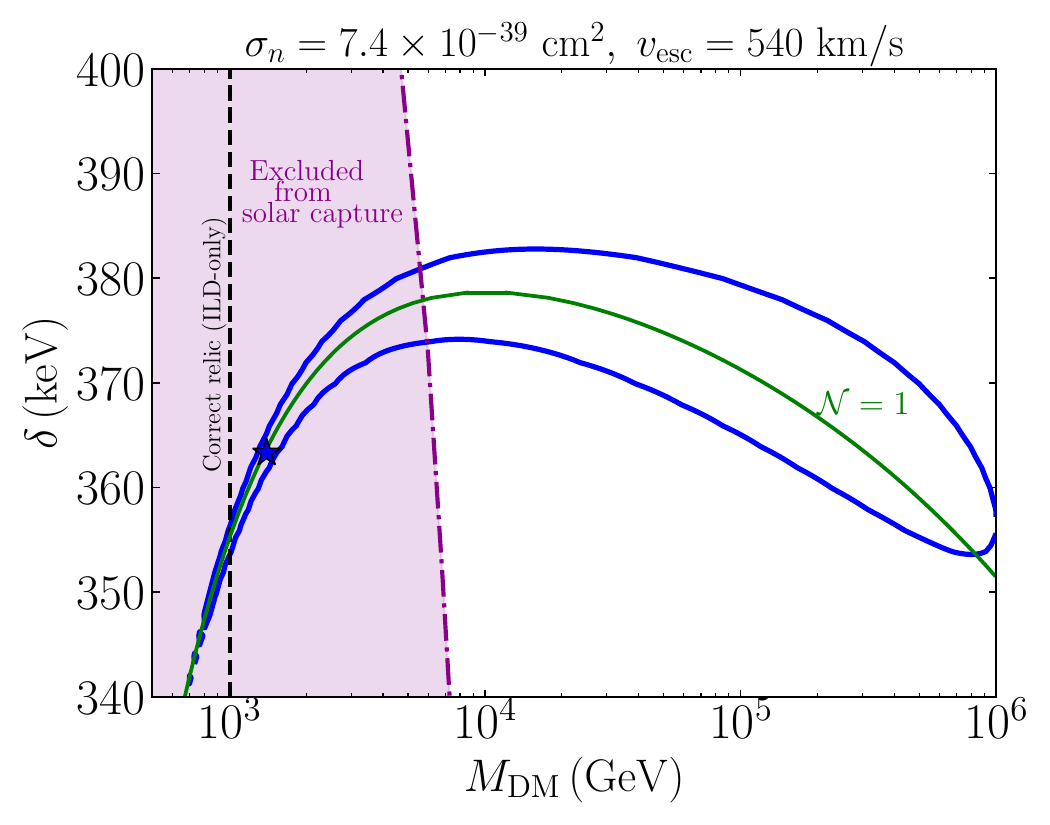}
\caption{$1\sigma$ region accommodating LZ230616 event in the plane of $\delta$ versus $M_{\rm DM}$ for the inelastic cross-section $\sigma_n=7.4\times10^{-39}{~\rm cm^2}$ is shown with blue contour. The best fit value ($\delta^{\rm BF}=363.32$ keV, $M^{\rm BF}_{\rm DM}=1400.66$ GeV) is shown with a blue start. The green contour represents $\mathcal{N}=1$. The dark magenta shaded region is ruled out from solar capture constraint in the limit of vanishing elastic scattering \cite{Langhoff:2026ujr}. The black dashed vertical line represents the correct relic contour for the minimal ILD DM case.}
\label{fig:lz1}
\end{figure}

In Fig.~\ref{fig:lz1}, we show the resulting $1\sigma$ confidence region accommodating the LZ230616 event in the $(M_{\rm DM},\delta)$ plane by fixing $\sigma_n=7.4\times10^{-39}~{\rm cm}^2$. The solid blue contour corresponds to the $1\sigma$ region, while the green contour represents the parameter space for which the expected number of signal events is exactly one. The dark magenta shaded region is excluded by the solar capture constraint. Here, we have assumed that the DM elastic cross-section is suppressed. This can naturally arise in a model with inert lepton doublet and triplet scalar, as will be discussed in Sec. \ref{sec:model}. The black dashed vertical line indicates the DM mass that gives the correct relic abundance in the minimal ILD DM scenario. It can be seen that the minimal ILD scenario is excluded by the solar capture constraint. Moreover, the minimal ILD DM remains overabundant for $M_{\rm DM}\geq 1.05$ TeV, and hence cannot account for the observed relic abundance in the high mass region. This situation can be resolved by extending the minimal ILD scenario with a scalar triplet. The presence of the triplet introduces additional annihilation channels, allowing the DM relic abundance to be brought down to the observed value even for larger DM masses, as discussed in Sec \ref{sec:model}.

The characteristic shape of the confidence region can be understood as follows. For smaller values of $M_{\rm DM}$, the allowed mass splitting $\delta$ increases with increasing $M_{\rm DM}$, reaching a maximum around $M_{\rm DM}\sim10^4$ GeV. As $M_{\rm DM}$ increases, a larger mass splitting can be accommodated while still allowing the required recoil energy to be produced by the high-velocity DM particles. However, for $M_{\rm DM}\gg m_N$, the reduced mass approaches the nuclear mass, and the kinematic dependence on $M_{\rm DM}$ becomes weak. At the same time, the DM number density decreases with increasing $M_{\rm DM}$, leading to a suppression of the scattering rate. Consequently, at larger $M_{\rm DM}$, smaller values of $\delta$ are preferred to compensate for the reduction in the event rate. This results in the characteristic turnover followed by a gradual decrease of $\delta$ at large $M_{\rm DM}$. The green $\mathcal{N}=1$ contour follows a similar trend. Thus, the scalar triplet extension provides a viable high-mass realization of the ILD scenario that can simultaneously account for the observed relic abundance and the LZ event.

\section{Model Realization}\label{sec:model}
In this section, we provide a detailed discussion on ILD DM model extended with a triplet scalar \cite{Bhattacharya:2018fus}. We extend the SM by a vector-like $SU(2)_L$ lepton doublet,
\begin{equation*}
L' = \begin{pmatrix} N' \\ E' \end{pmatrix}, \qquad
		SU(2)_L \ \text{doublet}, \quad Y = -\tfrac12.
\end{equation*}
Additionally, we impose a $Z_2$ symmetry, under which $L'$ is odd while all SM particles carry an even charge. This forbids any mixing between $L'$ with SM lepton sector. Thus, the neutral component $N'$ serves as the DM candidate in this setup. The Lagrangian of this minimal setup is given by:
\begin{align}\label{eq:LagILD}
\mathcal{L}_\text{ILD} \supseteq \bar L' i\gamma^\mu D_\mu L' - M \bar L' L',
\end{align}
where $D_\mu = \partial_\mu - i g_2 W_\mu^a \frac{\tau^a}{2} - i g_1 Y B_\mu$, and $M$ is the bare Dirac mass of the components of the ILD.

We then add a small Majorana mass $m_\Delta$ $(\mathcal{O}(\text{keV}))$ for the $N'$ (the origin of such a Majorana term is discussed later in this section, which is otherwise forbidden in this minimal setup by requiring gauge invariance). This results in a splitting of the neutral component into two pseudo-Dirac states $N'_1$ and $N'_2$ with masses
\begin{equation}
    M_{1}=M-m_\Delta(=M_\text{DM}) \quad M_2=M+m_\Delta,
\end{equation}
and the mass splitting between the two pseudo-Dirac states is $\delta=2m_\Delta$.
The interaction terms can be expanded in the physical basis as:
\begin{align}
\mathcal{L}_\gamma^{\rm NC} &= -e\, \bar E'\gamma^\mu E'\, A_\mu,\\
\mathcal{L}_Z^{\rm NC}&= \bar N'\gamma^\mu N'= \frac{g_2}{2\cos\theta_W}\, i\,\overline{N'_1}\gamma^\mu N'_2\, Z_\mu,\\
\mathcal{L}_W^\text{CC} &= \frac{g_2}{\sqrt2}\,\bar E'\gamma^\mu N' W^-_\mu + \text{h.c.} \nonumber\\
&= \frac{g_2}{2}\bar E'\gamma^\mu\left(N'_1+iN'_2\right) W^-_\mu + \text{h.c.}.
\end{align}

We further extend the SM with a scalar triplet $\Delta\sim(\mathbf{1},\mathbf{3},1)$, represented in the matrix notation as:
\begin{equation}
\Delta = \begin{pmatrix} \Delta^+/\sqrt{2} & \Delta^{++} \\ \Delta^0 & -\Delta^+/\sqrt{2} \end{pmatrix}.
\end{equation}
The relevant Lagrangian can be written as:
\begin{align}
\label{eq:Lagtriplet}
\mathcal{L}&\supseteq  \mathrm{Tr}\big[(D_\mu\Delta)^\dagger
		(D^\mu\Delta)\big]-\frac{1}{\sqrt2}(y_L)_{\alpha\beta}\Big[\overline{L_\alpha^c}(i\sigma_2\Delta)L_\beta+{\rm h.c.}\Big]\nonumber\\
&\quad-V(H,\Delta),
\end{align}
where $D_\mu = \partial_\mu - i g_2 W_\mu^a \frac{\tau^a}{2} - i g_1 Y B_\mu$. The first term corresponds to the kinetic term, whereas the second term is responsible for Majorana neutrino mass via the type-II seesaw mechanism. The scalar potential $V(H,\Delta)$ is given by:
\begin{align}
V(H,\Delta) &= -\mu_H^2 H^\dagger H+\lambda_H(H^\dagger H)^2+M_\Delta^2\,{\rm Tr}(\Delta^\dagger\Delta)\nonumber\\
&\quad+\lambda_\Delta\big[{\rm Tr}(\Delta^\dagger\Delta)\big]^2+\lambda_{\Delta2}\,{\rm Tr}\big[(\Delta^\dagger\Delta)^2\big]\nonumber\\
&\quad+\lambda_3(H^\dagger H)\,{\rm Tr}(\Delta^\dagger\Delta)+\lambda_4\,H^\dagger\Delta\Delta^\dagger H\nonumber\\
&\quad+\big(\mu\,H^\dagger\Delta\widetilde H+{\rm h.c.}\big),
\end{align}
where $H=(0\quad (h+v_h)/\sqrt{2})^T$ and $\widetilde H=i\sigma_2H^*$. After electroweak symmetry breaking, the scalar triplet acquires an induced VEV:
\begin{equation}
v_\Delta \approx -\frac{\mu v_h^2}{2 M_\Delta^2 + (\lambda_{3} + \lambda_{4}) v_h^2}.
\end{equation}
Thus, the mass splitting can be expressed as $\delta=2m_\Delta=2\sqrt{2}y_\Delta v_\Delta$
Similarly, the Majorana neutrino mass can be written as:
\begin{equation}
    \left(M_\nu\right)_{\alpha\beta}=\sqrt{2}\left(y_L\right)_{\alpha\beta}v_\Delta.
\end{equation}

In the minimal ILD setup \cite{Bhattacharya:2018fus}, DM annihilation and co-annihilation proceed purely through $W^\pm$ and $Z$ mediation, so the relic density is fixed entirely by these processes. As a result, the correct relic abundance is achieved only at a single point, $M_\text{DM} \simeq 1.05$~TeV -- below this mass the relic density is under-abundant, and above it, over-abundant. Moreover, the model's large elastic scattering cross-section runs afoul of direct-detection constraints, disfavoring it further. 
\begin{figure}[h]
\centering
\includegraphics[width=0.95\linewidth]{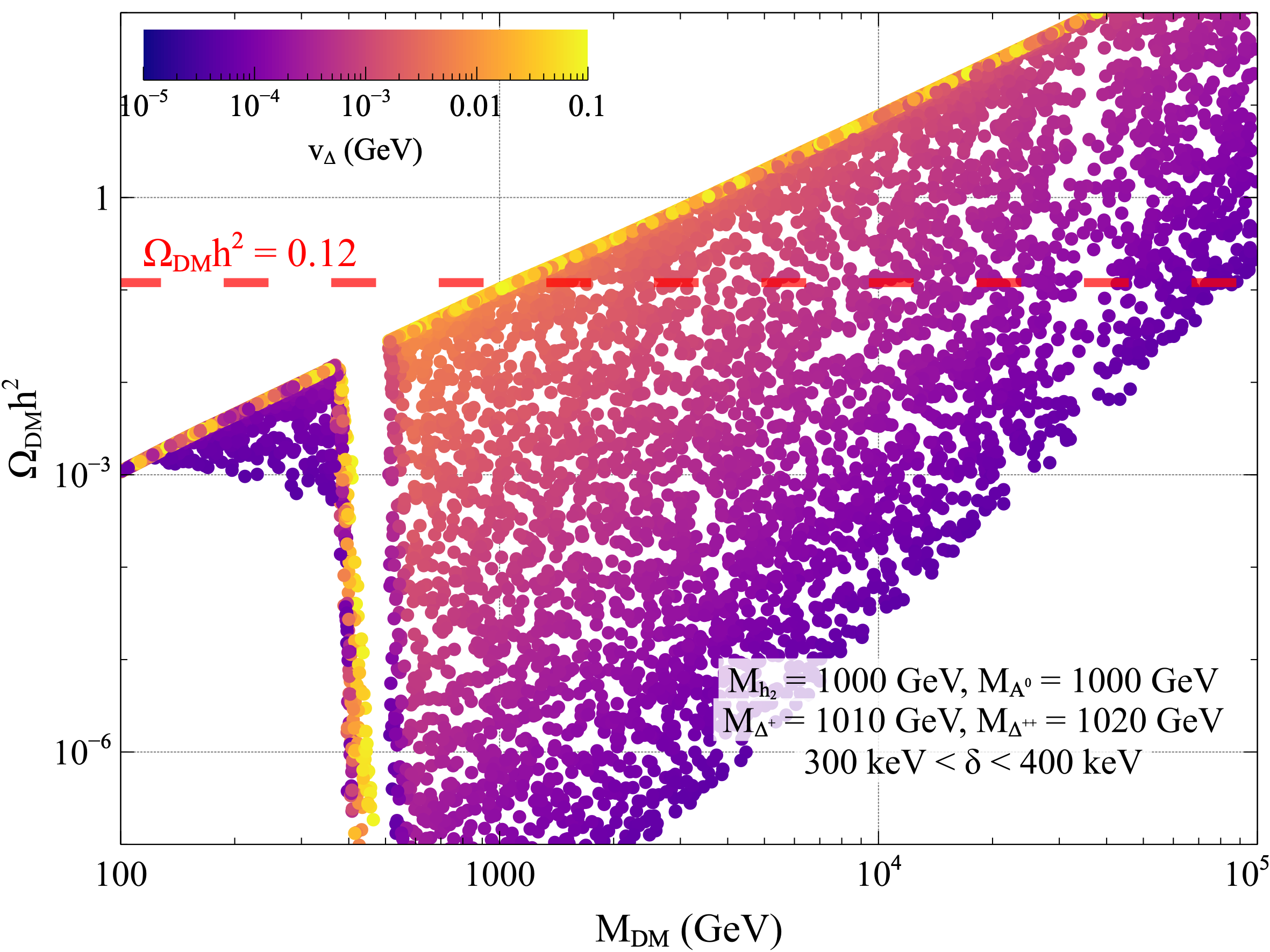}
\caption{DM relic density is shown as a function of DM mass. The colormap shows the value of $v_\Delta$. The red dashed lines corresponds to rewuired relic density, $\Omega_\text{DM}h^2=0.12$.}
\label{fig:relic01}
\end{figure}
In our setup, however, the presence of the scalar triplet opens up additional annihilation and co-annihilation channels mediated by $\Delta$, allowing the DM to remain in equilibrium for a longer epoch at higher masses. This, in turn, enables the correct relic density to be achieved even in the heavier DM mass region, which was otherwise over-abundant in the minimal setup. To estimate the DM relic abundance, we use \texttt{micrOMEGAs}~\cite{Alguero:2022inz} and present the relic density as a function of DM mass in Fig.~\ref{fig:relic01}, with the color map denoting the values of $v_\Delta$; the masses of the scalar triplet sector are held fixed, as indicated in the figure inset. We further impose a filter on the mass splitting, $300~\text{keV} < \delta < 400~\text{keV}$, as preferred for explaining the LZ event (see the discussion in Section~\ref{sec:LZevent}). As the DM mass increases, the annihilation cross-section decreases, an effect compensated by increasing $y_\Delta \, (= \delta/2\sqrt{2}v_\Delta)$ or decreasing $v_\Delta$, as is evident from Fig.~\ref{fig:relic01}. Moreover,  we also find that the small mass splitting $\delta$ has a negligible effect on the relic parameter space.
\begin{figure}[H]
\centering
\includegraphics[width=1\linewidth]{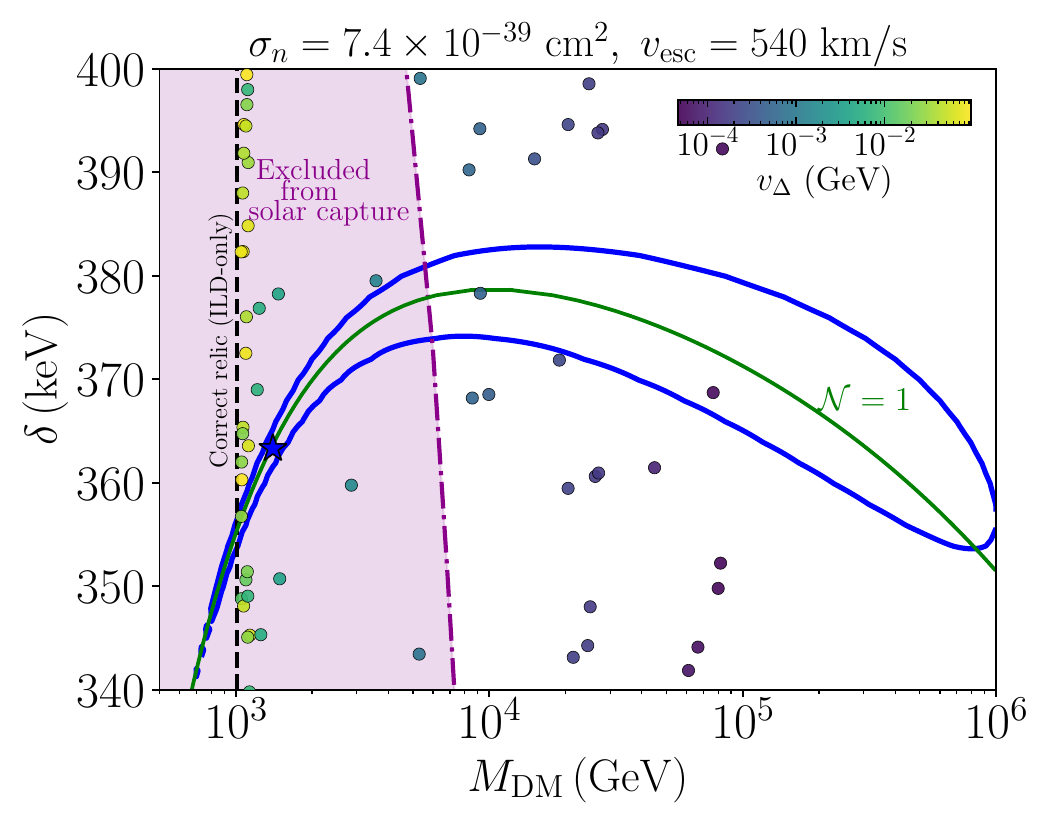}
\caption{Correct DM relic density satisfying points projected in the plane of $M_\text{DM}$ and $\delta$, where the color map shows the values of $v_\Delta$. The region left to the black dashed line remains the under-abundant. The other constraints are same as Fig. \ref{fig:lz1}.}
 \label{fig:relic02}
\end{figure}
In Fig.~\ref{fig:relic02}, we showcase the correct relic points in the plane of $\delta$ and $M_\text{DM}$, while the color map represents the values of $v_\Delta$. We also show the parameter space consistent with the 248 keV LZ event, same as Fig. \ref{fig:lz1}. The region left to the black dashed line remain under abundant. The correct relic density parameter space lies within the preferred region that accommodates the observed LZ230616 event while also evading the solar capture constraint.

\section{Conclusion}\label{sec:conclusion}
In this work, we explored the possibility of explaining the 248 keV recoil event observed at LZ in an ILD dark matter framework with masses larger than 1 TeV. Any electroweak DM with masses near 1 TeV faces severe constraints from solar capture. Moreover, DM masses $\gtrsim1$ TeV are over produced. We, thus, considered a minimal extension of the ILD DM with a scalar triplet which generates a small mass splitting among the neutral components of the ILD and introduces inelasticity in the framework. The addition of the scalar triplet also introduces additional annihilation processes through which the overabundant relic density of the ILD at larger DM masses can be brought down to the observed value. The triplet scalar can simultaneously address the non-zero neutrino mass via the type-II seesaw mechanism. This makes the framework minimal while establishing a connection between neutrino physics and DM, and provides a viable explanation of the 248 keV LZ event. This model can be easily extended to a vector-like fourth generation neutrino DM and sneutrino DM in supersymmetric scenario.

\section*{Acknowledgment}
P.K.P. acknowledges the Ministry of Education, Government of India, for providing financial support for his research via the Prime Minister’s Research Fellowship (PMRF) scheme. S.K.S acknowledges the support provided by IMSc, Chennai, during his visit.

\appendix
\section{Scalar sector mass spectrum}
In the presence of trilinear coupling in Eq.~\eqref{eq:Lagtriplet} between $\Delta$ and SM Higgs induces mass mixing among the CP-even components. The corresponding mass squared matrix $\mathcal{M}^2$ is given as:
\begin{equation}
\mathcal{M}^2 = \begin{pmatrix} M_{11}^2 & M_{12}^2 \\
M_{12}^2 & M_{22}^2 \end{pmatrix},
\end{equation}
where
\begin{align}
M_{11}^2 &= 2\lambda_H v^2, \\
M_{22}^2 &= -\frac{\mu v^2}{\sqrt2\,v_\Delta} + 2(\lambda_\Delta{+}\lambda_{\Delta2})v_\Delta^2, \\
M_{12}^2 &= v\big[(\lambda_3{+}\lambda_4)v_\Delta + \sqrt2\,\mu\big]. 
\end{align}
Diagonalizing in the physical basis $(h_1,h_2)$
\begin{equation}
\begin{pmatrix} h_1 \\ h_2 \end{pmatrix}
= \begin{pmatrix}\cos\alpha & -\sin\alpha \\
\sin\alpha & \cos\alpha\end{pmatrix}
\begin{pmatrix}h \\ \delta'\end{pmatrix},
\end{equation}
where $h$ and $\delta'$ are the fluctuations of the SM Higgs and CP-even scalar triplet, respectively. $\alpha$ represents the mixing angle between the two CP-even scalar states and is given by,
\begin{equation}
\tan2\alpha = \frac{2\mathcal{M}^2_{12}}
{\mathcal{M}^2_{22}-\mathcal{M}^2_{11}},
\end{equation}
\begin{align}
m^2_{h_1,h_2} = &\frac{\mathcal{M}^2_{11}+\mathcal{M}^2_{22}}{2}\nonumber\\
&\mp \sqrt{\left(\frac{\mathcal{M}^2_{11}
-\mathcal{M}^2_{22}}{2}\right)^2
+ \left(\mathcal{M}^2_{12}\right)^2} .
\end{align}
$h_1\to h$ as $\alpha\to0$ is identified with the SM Higgs; $h_2\to\delta'$ is the heavy, mostly-triplet scalar.

%

\end{document}